\documentstyle[12pt]{article}

\font\twlgot =eufm10 scaled \magstep1
\font\egtgot =eufm8
\font\sevgot =eufm7

\font\twlmsb =msbm10 scaled \magstep1
\font\egtmsb =msbm8
\font\sevmsb =msbm7

\newfam\gotfam

\textfont\gotfam\twlgot
\scriptfont\gotfam\egtgot
\scriptscriptfont\gotfam\sevgot

\newfam\msbfam
\textfont\msbfam\twlmsb
\scriptfont\msbfam\egtmsb
\scriptscriptfont\msbfam\sevmsb
\def\Bbb{\protect\pBbb}
\def\pBbb{\relax\ifmmode\expandafter\Bb\else\typeout{You cann't use
Bbb in text mode}\fi}
\def\Bb #1{{\fam\msbfam\relax#1}}

\def\op#1{\mathop{{\it\fam0} #1}\limits}

\newcommand{\im}{{\rm Im\, }}
\newcommand{\re}{{\rm Re\, }}

\newcommand{\lng}{\langle}
\newcommand{\rng}{\rangle}

\newcommand{\intr}{{\rm Int\,}}

\newcommand{\bite}{\begin{itemize}}
\newcommand{\eite}{\end{itemize}}
\newcommand{\benu}{\begin{enumerate}}
\newcommand{\eenu}{\end{enumerate}}
\newcommand{\bde}{\begin{description}}
\newcommand{\ede}{\end{description}}
\newcommand{\bquo}{\begin{quote}}
\newcommand{\equo}{\end{quote}}
\newcommand{\bquot}{\begin{quotation}}
\newcommand{\equot}{\end{quotation}}

\newcommand{\eqref}[1]{(\ref{#1})}
\newcommand{\beq}{\begin{equation}}
\newcommand{\eeq}{\end{equation}}
\newcommand{\ben}{\begin{eqnarray}}
\newcommand{\een}{\end{eqnarray}}
\newcommand{\be}{\begin{eqnarray*}}
\newcommand{\ee}{\end{eqnarray*}}
\newcommand{\bea}{\begin{eqalph}}
\newcommand{\eea}{\end{eqalph}}

\newcommand{\cE}{{\cal E}}

\newcommand{\al}{\alpha}

\newcommand{\dl}{\delta}
\newcommand{\la}{\lambda}

\newcommand{\f}{\phi}
\newcommand{\vf}{\varphi}

\newcommand{\G}{\Gamma}

\newcommand{\th}{\theta}

\newcommand{\wt}{\widetilde}
\newcommand{\wh}{\widehat}
\newcommand{\ol}{\overline}

\newcommand{\dr}{\partial}

\let\ssection=\section
\renewcommand{\section}{\setcounter{equation}{0}\ssection}

\newcounter{eqalph}
\newcounter{equationa}
\newcounter{remark}
\newcounter{example}
\newcounter{theorem}
\newcounter{proposition}
\newcounter{lemma}
\newcounter{corollary}
\newcounter{definition}
\newenvironment{eqalph}{\stepcounter{equation}
\setcounter{equationa}{\value{equation}}
\setcounter{equation}{0}

\begin{eqnarray}}{\end{eqnarray}\setcounter{equation}{\value{equationa}}}

\def\theremark{\arabic{remark}}
\def\therexample{\arabic{remark}}

\def\thedefinition{\arabic{definition}}

\newenvironment{rem}{\refstepcounter{remark}\medskip\noindent{\it Remark
\theremark:}}{\medskip}

\newenvironment{theo}{\refstepcounter{definition} \medskip
\noindent{\it Theorem \thedefinition:}}{\medskip}

\newcommand{\mar}[1]{}

\begin{document}
\hbox{}

{\parindent=0pt

{\large\bf On the mathematical hypothesis of phenomena like the
confinement}
\bigskip

{\sc Gennadi Sardanashvily}

{\sl 
Department of Theoretical Physics, Moscow State
University, Russia}
\bigskip
\bigskip

{\small

The Wick rotation provides the standard technique of computing Feynman
diagrams by means of Euclidean propagators. Let us suppose that quantum
fields in an interaction zone are really Euclidean. In contrast with the
well-known Euclidean field theory dealing with the Wightman and Schwinger
functions of free fields, we address complete Green's functions of
interacting fields, i.e., causal forms on the Borchers algebra of
quantum fields. They are the Laplace transform of the Euclidean
states obeying a certain condition. If Euclidean states of 
a quantum field system,
e.g., quarks  do not satisfy this
condition, this system fails to possess Green's functions and,
consequently, the
$S$-matrix. One therefore may conclude that it is not observed in the
Minkowski space.

}

}

\bigskip

\bigskip

As is well known, the Wick rotation enables one to compute the Feynman
diagrams of perturbative quantum field theory by means of Euclidean
propagators. Let us suppose that it is not a technical trick, but quantum
fields in an interaction zone are really Euclidean.

For the sake of simplicity, we here restrict our consideration to real
scalar fields. One associates to them the Borchers algebra
\mar{qm801}\beq
A_{RS^4}=\Bbb R\oplus RS^4\oplus RS^8\oplus\cdots, \label{qm801}
\eeq
where $RS^4$ is the nuclear space of smooth real functions of rapid
decrease on $\Bbb R^n$ \cite{borch,hor}. It is the real subspace of 
the space $S(\Bbb R^4)$
of smooth complex functions of rapid decrease on $\Bbb R^4$. Its 
topological dual is the space
$S'(\Bbb R^4)$ of tempered distributions (generalized functions)
\cite{piet,bog}.
Any continuous positive form on the Borchers algebra $A_{RS^4}$
(\ref{qm801}) is represented 
by a collection
of tempered distributions $\{W_n\in S'(\Bbb R^{4n})\}$ such that
\mar{qm810}\beq
f(\psi_n)=
\int W_n(x_1,\ldots,x_n)\psi_n(x_1,\ldots,x_n)d^4x_1\cdots d^4x_n,
\qquad \psi_n\in RS^{4n}. \label{qm810}
\eeq
For instance, these are the Wightman functions describing free scalar
quantum fields in the Minkowski space. 

We address the causal forms $f^c$ on the Borchers algebra 
$A_{RS^4}$ (\ref{qm801}) characterizing quantum fields
created at some instant
and annihilated at another one. 
They are given by the functionals
\mar{1260,030}\ben
&&f^c(\psi_n)=
\int W_n^c(x_1,\ldots,x_n)\psi_n(x_1,\ldots,x_n)d^4x_1\cdots d^4x_n,
\qquad \psi_n\in RS^{4n}, \label{1260}\\
&& W^c_n(x_1,\ldots,x_n)=
\op\sum_{(i_1\ldots i_n)}\th(x^0_{i_1}-x^0_{i_2})
\cdots\th(x^0_{i_{n-1}}-x^0_{i_n})W_n(x_1,\ldots,x_n), \label{030}
\een
where $W_n\in S'(\Bbb R^{4n})$ are tempered distributions, $\th$ is the step
function, and the sum runs
through all permutations $(i_1\ldots i_n)$ of the tuple of
numbers $1,\ldots,n$ \cite{bog2}.
A problem is that the functionals $W^c_n$ (\ref{030}) need not
be tempered distributions. For instance, $W^c_1\in S'(\Bbb R)$ iff
$W_1\in S'(\Bbb R_\infty)$, where $\Bbb R_\infty$ is the compactification
of $\Bbb R$ by means of the point $\{+\infty\}=\{-\infty\}$. Moreover,
the causal forms are not positive. Therefore, they do not provide states
of the  Borchers algebra $A_{RS^4}$
in general. At the same time, the causal forms come
from the Laplace transformation of Euclidean states of the
Borchers algebra \cite{sard91,sard02}. These Euclidean states however
are not arbitrary but must take the form (\ref{036}). If Euclidean 
states of 
a quantum field system,
e.g., quarks  do not satisfy this
condition, this system fails to possess Green's functions and,
consequently, the
$S$-matrix. One therefore may conclude that it is not observed in the
Minkowski space.

Note that, since the causal forms (\ref{030}) are symmetric,
the Euclidean states of the Borchers algebra $A_{RS^4}$
can be obtained as states of the corresponding commutative tensor
algebra $B_{RS^4}$ \cite{sard91,sard02}. Provided with the
direct sum topology,
$B_{RS^4}$ becomes a topological
involutive algebra. It coincides with
the enveloping algebra of the Lie algebra
of the additive Lie group $T_{RS^4}$ of translations in
$RS^4$. Therefore, one can obtain the states of the algebra
$B_{RS^4}$ by constructing cyclic strongly continuous unitary
representations of the nuclear Abelian group $T_{RS^4}$.
Such a representation is
characterized by a positive-definite continuous generating
function $Z$ on $RS^4$
which is the Fourier
transform 
of a bounded positive measure of total mass 1
on the space  $S'(\Bbb R^4)$ of generalized functions \cite{gelf64}. This
generating function $Z$ plays the role of a generating functional of
Euclidean Green's functions represented by a functional integral on
$S'(\Bbb R^4)$.

It should be emphasized that the above mentioned Euclidean states differ
from the well-known Schwinger functions in the so called Euclidean field
theory \cite{bog,sim74,zin}. They are the Laplace transform of Wightman
functions, and describe free Euclidean quantum fields (see Appendix).
Note that the Euclidean counterpart of time ordered correlation functions
is also considered in the Euclidean quantum field theory, but not by
means of the Wick rotation \cite{guer}.

\bigskip
\bigskip

In order to describe the Wick rotation of Euclidean states, 
we start with
the basic formulas of the Fourier--Laplace (henceforth FL) transformation
\cite{bog}. It 
is defined on Schwartz distributions, but
we focus on the tempered ones.

Recall that
 functions of rapid decrease on $\Bbb R^n$
are  complex smooth functions $\psi(x)$ such that
the quantities
\mar{spr453}\beq
\|\psi\|_{k,m}=\op\max_{|\al|\leq k} \op\sup_x(1+x^2)^m|D^\al \psi(x)|
\label{spr453}
\eeq
are finite for all $k,m\in \Bbb N$. Here, we follow 
the standard notation
\be
D^\al=\frac{\dr^{|\al|}}{\dr^{\al_1} x^1\cdots\dr^{\al_n}x^n}, \qquad
|\al|=\al_1+\cdots +\al_n, 
\ee
for an $n$-tuple of natural numbers $\al=(\al_1,\ldots,\al_n)$.
The functions of rapid decrease constitute the nuclear space
$S(\Bbb R^n)$ with respect to the
topology determined by the seminorms (\ref{spr453}). Its dual
is the space $S'(\Bbb R^n)$ of tempered distributions. The
corresponding contraction form is written as
\be
\lng \psi,h\rng=\op\int \psi(x) h(x) d^nx, \qquad \psi\in S(\Bbb R^n),
\qquad h\in S'(\Bbb R^n).
\ee
The space
$S(\Bbb R^n)$ is provided with the nondegenerate
separately continuous Hermitian form
\be
\lng \psi|\psi\rng=\int \psi(x)\ol{\psi(x)}d^nx. 
\ee
The completion of $S(\Bbb R^n)$
with respect to this form is the space $L^2_C(\Bbb R^n)$ of
square integrable complex functions on $\Bbb R^n$.
We have the rigged Hilbert space
\be
S(\Bbb R^n)\subset L^2_C(\Bbb R^n) \subset S'(\Bbb R^n).
\ee
Let $\Bbb R_n$ denote the dual of $\Bbb R^n$ coordinated by
$(p_\la)$. The Fourier transform
\mar{spr460,1}\ben
&& \psi^F(p)=\int \psi(x)e^{ipx}d^nx, \qquad px=p_\la x^\la,
\label{spr460}\\
&& \psi(x)=\int \psi^F(p)e^{-ipx}d_np, \qquad d_np=(2\pi)^{-n}d^np,
\label{spr461}
\een
provides an isomorphism between the spaces $S(\Bbb R^n)$ and
$S(\Bbb R_n)$.
The Fourier transform of tempered distributions is defined by the
condition
\be
\int h(x)\psi(x)d^nx=\int h^F(p)\psi^F(-p)d_np,
\ee
and is written in the form
(\ref{spr460}) -- (\ref{spr461}).
It provides an isomorphism between the spaces of tempered distributions
$S'(\Bbb R^n)$ and $S'(\Bbb R_n)$.

Let $\Bbb R^n_+$ and $\ol\Bbb R^n_+$ further denote the subset of points
of $\Bbb R^n$ with strictly positive Cartesian coordinates
and its closure, respectively. 
Let $f\in S'(\Bbb R^n)$ be a tempered distribution and $\G(f)$ the 
convex subset
of points $q\in \Bbb R_n$ such that
\mar{1273}\beq
e^{-qx}f(x)\in S'(\Bbb R^n). \label{1273}
\eeq
In particular, $0\in\G(f)$. Let $\intr\G(f)$ and $\dr\G(f)$ denote
the interior and boundary of $\G(f)$, respectively.
The FL transform 
of a tempered distribution $f\in S'(\Bbb R^n)$ is defined as the
tempered distribution
\mar{7.2}\beq
f^{FL}(p+iq)=(e^{-qx}f(x))^F(p)=\int f(x)e^{i(p+iq)x} d^nx \in S'(\Bbb R_n),
\label{7.2}
\eeq
which is the Fourier transform of the distribution (\ref{1273}) depending on
$q$ as parameters. One can think of the FL 
transform (\ref{7.2})
as being the Fourier transform with respect to the complex arguments $k=p+iq$.

If $\intr\G(f)\neq\emptyset$, the FL transform 
$f^{FL}(k)$ is a holomorphic
function $h(k)$ of complex arguments $k=p+iq$ on the open tube
$\Bbb R_n +i\intr\G(f)\subset \Bbb C_n$
over $\intr\G(f)$. Moreover, for any compact subset $Q\subset \intr\G(f)$,
there exist strictly positive numbers $A$ and $m$, depending on $Q$ and
$f$, such that
\mar{7.3}\beq
|f^{FL}(p+iq)|\leq A(1+|p|)^m, \qquad p\in \Bbb R_n, \qquad q\in Q. \label{7.3}
\eeq
The evaluation (\ref{7.3}) is 
equivalent to the
fact that the function $h(p+iq)$ defines a family of tempered distributions
$h_q(p)\in S'(\Bbb R_n)$ of the variables $p$ depending continuously on
parameters $q\in S$. If $0\in\intr\G(f)$, then
\be
f^{FL}(p+i0)=\op\lim_{q\to 0}f^{FL}(p+iq)
\ee
coincides with the Fourier transform $f^F(p)$ of $f$. The case of
$0\not\in\intr\G(f)$ is more intricate. Let $S$ be a convex domain in 
$\Bbb R^n$
such that $0\in\dr S$, and let $h(p+iq)$ be a holomorphic function on the tube
$\Bbb R_n +iS$ which defines a family of tempered distributions
$h_q(p)\in  S'(\Bbb R_n)$,
depending on parameters $q$. One says that $h(p+iq)$ has a 
generalized boundary
value $h_0(p)\in S'(\Bbb R_n)$ if, for any
frustum $K^r\subset S\cup \{0\}$ of the cone $K\subset \Bbb R_n$ (i.e.,
$K^r=\{q\in K\,:\,|q|\leq r\}$), one has
\be
h_0(\psi(p))=\op\lim_{|q|\to 0,\,q\in K^r\setminus\{0\}} h_q(\psi(p)) 
\ee
for all functions $\psi\in S(\Bbb R_n)$ of rapid decrease. Then the following
holds \cite{bog}.

\begin{theo}
Let $f\in S'(\Bbb R^n)$, $\intr\G(f)\neq \emptyset$ and $0\not\in\intr\G(f)$.
A generalized boundary value of the FL transform $f^{FL}(k)$
in $S'(\Bbb R_n)$ exists and coincides with the Fourier transform $f^F(p)$ of
the distribution $f$.
\end{theo}

Let us apply this result to the following important case.
The support of a tempered distribution
Ê$f$ is defined as the complement
of the maximal open subset $U$ where $f$ vanishes, i.e., $f(\psi)=0$ for all
$\psi\in S(\Bbb R^n)$ of support in $U$. Let $f\in S'(\Bbb R^n)$ be 
of support in
$\ol\Bbb R_+^n$. Then $\ol\Bbb R_{n+}\subset \G(f)$, and the 
FL transform
$f^{FL}$ is a holomorphic function on the tube over $\Bbb R_{n+}$, 
while its generalized
boundary value in $S'(\Bbb R_n)$ is given by the equality
\be
h_0(\psi(p))=\op\lim_{|q|\to 0,\,q\in\Bbb R_{n+}} 
f^{FL}_q(\psi(p))=f^F(\psi(p)), \qquad \forall
\psi\in S(\Bbb R_n).
\ee
Conversely, one can restore a tempered distribution $f$ of
support in $\ol\Bbb R_+^n$ from its FL transform
$h(k)=f^{FL}(k)$ even if this function is known only on
$i\Bbb R_{n+}$. Indeed, the formulas
\mar{7.4,5}\ben
&& \wt h(\phi)=\op\int_{\Bbb R_{n+}}h(iq)\f(q)d_nq= \op\int_{\Bbb R_{n+}}
d_nq \op\int_{\ol\Bbb R^n_+} e^{-qx}f(x)\f(q)d^nx= \label{7.4}\\
&& \qquad \op\int_{\ol\Bbb R^n_+} f(x)\wh\f(x)d^nx, \qquad \f\in S(\Bbb
R_{n+}),\nonumber\\
&& \wh\f(x)=\op\int_{\Bbb R_{n+}} e^{-qx}\f(q)d_nq, \qquad
x\in\ol\Bbb R^n_+, \qquad \wh\f\in S(\ol\Bbb R^n_+), \label{7.5}
\een
define a linear continuous functional $\wt h(q)=h(iq)$ on the
space $S(\Bbb R_{n+})$. It is called the Laplace transform
$f^L(q)=f^{FL}(iq)$ of a tempered distribution $f$.

\begin{rem}
Let us illustrate the restoration of a tempered distribution from the
functional (\ref{7.4}) in the case of $n=1$. Let $f\in S'(\ol\Bbb R_+)$.
Its FL transform reads
\mar{1275}\beq
h(p+iq)=\op\int^\infty_0 e^{i(p+iq)x}f(x)dx, \qquad q>0. \label{1275}
\eeq
Since $h(p+iq)$ (\ref{1275}) has the generalized boundary value
$h(p+i0)$, the $f$ is reconstructed from $\wh h(z)=h(iz)$ by the formulas
\mar{7.6}\beq
f(x)=\frac{1}{2\pi i}\op\int^{0+i\infty}_{0-i\infty}e^{zx}\wt h(z)dz,
\qquad \wt h(0-ip)=f^F(p), \label{7.6}
\eeq
where
\be
\wt h(q)=h(iq)=\op\int^\infty_0e^{-qx}f(x)dx, \qquad q>0,
\ee
is the Laplace transform $f^L$ of $f$.
\end{rem}

The image of the space $S(\Bbb R_{n+})$ with respect to the mapping
$\f(q)\mapsto \wh \f(x)$ (\ref{7.5}) is dense in $S(\ol\Bbb R^n_+)$. 
Then the family
of seminorms $\|\f\|'_{k,m}=\|\wh\f\|_{k,m}$, where $\|.\|_{k,m}$ are seminorms
(\ref{spr453}) on $S(\Bbb R^n)$, determines the new coarsen 
topology on $S(\Bbb R_{n+})$ such that the functional (\ref{7.4}) 
remains continuous
with respect to this topology. Then the following is proved
\cite{bog}.

\begin{theo} \label{1276} \mar{1276}
The mappings (\ref{7.4}) and (\ref{7.5})
provide one-to-one correspondence between the Laplace transforms 
$f^L(q)=f^{FL}(iq)$
of tempered distributions $f\in S'(\ol\Bbb R^n_+)$ and the elements 
of $S'(\Bbb R_{n+})$
which are continuous with respect to the coarsen topology on $S(\Bbb R_{n+})$.
\end{theo}

\bigskip
\bigskip

With Theorem \ref{1276}, the above mentioned Wick rotation of
Green's functions of Euclidean quantum fields to
causal forms in the Minkowski space is described as follows.

Let us denote by $X$ the space $\Bbb R^4$ associated to the
real subspace of $\Bbb C^4$ and by $Y$ the space $\Bbb R^4$, coordinated by
$(y^0,y^{1,2,3})$ and associated to the subspace $\wt Y$ of $\Bbb C^4$
whose points possess the coordinates $(iy^0,y^{1,2,3})$. If $X$ is 
the Minkowski
space, then one can think of $Y$ as
being its Euclidean partner. 
Since $X$ and $Y$ in
$\Bbb C^4$ have the same spatial subspace, we
further omit the dependence on spatial coordinates.
Therefore, let us consider the complex plane $\Bbb C^1=X\oplus iZ$ of the time
$x$ and the Euclidean time $z$ and the complex plane $\Bbb C_1=P\oplus iQ$
of the associated momentum coordinates $p$ and $q$.

Let $W(q)\in S'(Q)$ be a tempered distribution such that
\mar{036}\beq
W=W_++W_-,\qquad W_+\in S'(\ol Q_+), \qquad W_-\in S'(\ol Q_-).\label{036}
\eeq
For instance, $W(q)$ is an ordinary function at $0$.
For every test function $\psi_+\in S(X_+)$, we have
\mar{037}\ben
&&\frac{1}{2\pi}\op\int_{\ol Q_+}W(q)\wh\psi_+(q)dq
=\frac{1}{2\pi}\op\int_{\ol
Q_+}dq\op\int_{X_+}dx [W(q)\exp(-qx)\psi_+(x)]=\nonumber\\
&&\qquad\frac{1}{(2\pi)^2}\op\int_{\ol Q_+}dq\op\int_Pdp\op
\int_{X_+}dx[W(q)\psi^F_+(p)\exp(-ipx-qx)]=\nonumber\\
&&\qquad \frac{-i}{(2\pi)^2}\op\int_{\ol Q_+}dq
\op\int_Pdp[W(q)\frac{\psi^F_+(p)}{p-iq}]=\frac{1}{2\pi}
\op\int_{\ol Q_+}W(q)\psi^{FL}_+(iq)dq, \label{037}
\een
due to the fact that the FL transform
$\psi^{FL}_+(p+iq)$ of the function $\psi_+\in S(X_+)\subset S'(X_+)$
exists and that it is holomorphic on the tube $P+iQ_+, Q_+$.
Moreover, $\psi^{FL}_+(p+i0)=\psi^F_+(p)$, and the function
$\wh\psi_+(q)=\psi^{FL}_+(iq)$ can be regarded as the Wick rotation
of the test function
$\psi_+(x)$. The equality (\ref{037}) can be brought into the form
\mar{038}\ben
&&\frac{1}{2\pi}\op\int_{\ol Q_+}W(q)\wh\psi_+(q)dq=
\op\int_{X_+}\wh W_+(x)\psi_+(x)dx, \label{038}\\
&&\wh W_+(x)=\frac{1}{2\pi}\op\int_{\ol Q_+}\exp(-qx)W(q)dq, \qquad
x\in X_+. \nonumber
\een
By virtue of Theorem  \ref{1276}, it associates to a distribution
$W(q)\in S'(Q)$ the distribution
$\wh W_+(x)\in S'(X_+)$, continuous with respect to
the coarsen topology on $S(X_+)$.

For every test function $\psi_-\in S(X_-)$, the similar relations
\mar{039}\ben
&&\frac{1}{2\pi}\op\int_{\ol Q_-}W(q)\wh\psi_-(q)dq=
\op\int_{X_-}\wh W_-(x)\psi_-(x)dx, \label{039}\\
&&\wh W_-(x)=\frac{1}{2\pi}\op\int_{\ol Q_-}\exp(-qx)W(q)dq, \qquad 
x\in X_-, \nonumber
\een
hold. Combining (\ref{038}) and (\ref{039}), we obtain
\mar{040}\ben
&&\frac{1}{2\pi}\op\int_QW(q)\wh\psi(q)dq=\op\int_X\wh W(x)\psi(x)dx, 
\label{040}\\
&&\wh\psi=\wh\psi_++\wh\psi_-, \qquad \psi=\psi_++\psi_-, \nonumber
\een
where $\wh W(x)$ is a linear functional on functions $\psi\in S(X)$, which
together with all derivatives vanish
at $x=0$. One can think of
$\wh W(x)$ as being the Wick rotation of the
distribution (\ref{036}). One should additionally define $\wh W$ at the point
$x=0$ in order to make it to a functional on the whole space $S(X)$. This
is the well-known ambiguity of chronological forms in quantum field theory.

In particular, let 
a tempered distribution 
\mar{1270}\beq
M(\f_1,\f_2)=\int W_2(x_1,x_2)\f_1(x_1)\f_2(x_2) \label{1270}
d^nx_1d^nx_2.
\eeq
be 
the Green's function
of some positive elliptic differential operator $\cE$, i.e.,
\be
\cE_{y_1}W_2(y_1,y_2)=\dl(y_1-y_2),
\ee
where $\dl$ is Dirac's $\dl$-function. Then the distribution $W_2$ reads
\mar{1272}\beq
W_2(y_1,y_2)=w(y_1-y_2), \label{1272}
\eeq
and we obtain the form
\be
&& F_2(\f_1\f_2)=M(\f_1,\f_2)=\int w(y_1-y_2)\f_1(y_1)\f_2(y_2) 
d^4y_1 d^4y_2=\\
&& \qquad \int w(y)\f_1(y_1)\f_2(y_1-y)d^4y d^4y_1=\int w(y)\vf(y)d^4y=
\int w^F(q)\vf^F(-q) d_4q, \\
&& y=y_1-y_2, \qquad \vf(y)=\int \f_1(y_1)\f_2(y_1-y)d^4y_1.
\ee
For instance, if
\be
\cE_{y_1} =-\Delta_{y_1}+m^2,
\ee
where $\Delta$ is the Laplacian, then
\mar{1271}\beq
w(y_1-y_2)=\int\frac{\exp(-iq(y_1-y_2))}{q^2+m^2}d_4q, \label{1271}
\eeq
where $q^2$ is the Euclidean square, is the propagator
of a massive Euclidean scalar field.
Let the Fourier transform $w^F$ of the distribution $w$ (\ref{1272})
satisfy the condition 
(\ref{036}). Then its Wick rotation (\ref{040}) is the functional
\be
\wh w(x)=\th(x)\op\int_{\ol Q_+}w^F(q)\exp(-qx)dq +
\th(-x)\op\int_{\ol Q_-}w^F(q)\exp(-qx)dq
\ee
on scalar fields in the Minkowski space. For instance, let $w(y)$ be the
Euclidean propagator
(\ref{1271}) of a massive scalar field. Then due to the analyticity of
\be
w^F(q)=(q^2+m^2)^{-1}
\ee
on the domain $\im q\cdot \re q>0$, one can show that
$\wh w(x)=-iD^c(x)$ 
where $D^c(x)$ is a familiar causal Green's function.

\bigskip
\bigskip

\noindent

{\bf Appendix}

\bigskip

Let us apply Theorem \ref{1276} to the relation between 
the Wightman and Schwinger functions in the Euclidean field theory in
order to show the difference between Schwinger functions and the above
mentioned states of Euclidean fields.

Recall that Wightman functions are defined as
tempered distributions $W_n\subset S'(\Bbb R^{4n})$
in the Minkowski space which obey the Garding--Wightman axioms
of axiomatic field theory \cite{bog,sim74,zin}.
Let us mention
the Poincar\'e covariance axiom, the spectrum condition and the 
locality condition. 
Due to the translation covariance of Wightman functions $W_n$,
there exist tempered distributions
$w_n\in S'(\Bbb R^{4n-4})$ such that
\mar{1278}\beq
W_n(x_1,\ldots,x_n)= w_n(x_1-x_2,\ldots,x_{n-1}-x_n). \label{1278}
\eeq
The spectrum condition implies that the Fourier transform $w^F_n$ of
the distributions $w_n$ (\ref{1278}) is of
support in
ÊÊÊÊthe closed forward light cone $\ol V_+$ in the momentum Minkowski
space $\Bbb R_4$. It follows that
the Wightman function $w_n$ is a generalized boundary value in 
$S'(\Bbb R^{4n-4})$
of the function $(w^F_n)^{FL}$, which is the FL transform
of the function $w^F_n$ with respect to variables $p^i_0$ and which is
holomorphic on the tube $(\Bbb R^4+iV_-)^{n-1}\subset \Bbb C^{4n}$. 
Accordingly,
$W_n(x_1,\ldots,x_n)$ is a generalized boundary value in $S'(\Bbb R^{4n})$
of a function $W_n(z_1,\ldots,z_n)$, holomorphic on the tube
\be
\{z_i\,:\, \im(z_{i+1}-z_i)\in V_-, \,\, \re z_i\in\Bbb R^4\}.
\ee
In accordance with the Lorentz covariance, the Wightman functions admit
an analytic continuation
onto a wider domain in $\Bbb C^{4n}$, called the extended forward tube.
Furthermore, the locality condition implies
that they are symmetric on this domain.

Let $X$ and $Y$ be 
the Minkowski
subspace and its Euclidean partner in
$\Bbb C^4$, respectively.
Let us consider the subset $\wt Y^n_{\neq}\subset \wt Y\subset \Bbb 
C^{4n}$ which consists
of the points $(z_1,\ldots,z_n)$ such that $z_i\neq z_j$. It belongs to
the domain of analyticity of the Wightman function $W_n(z_1,\ldots,z_n)$, 
whose restriction
to $\wt Y^n_{\neq}$ defines the symmetric function
\be
S_n(y_1,\ldots,y_n)=W_n(z_1,\ldots,z_n), \qquad z_i=(iy^0_i,y^{1,2,3}_i),
\ee
on $Y^n_{\neq}$. It is called the Schwinger function. 
On the domain $Y^n_<$ of points $(y_1,\ldots,y_n)$ such that
$0<y_1^0<\cdots< y_n^0$, the Schwinger function takes the form
\mar{1279}\beq
S_n(y_1,\ldots,y_n)=s_n(y_1-y_2,\ldots,y_{n-1}-y_n), \label{1279}
\eeq
where $s_n$ is an element of the space $S'(Y_-^{n-1})$ which is continuous
with respect to the coarsen topology on
$S(Y_-^{n-1})$. Consequently, in accordance with Theorem \ref{1276} 
and by virtue of the formula
(\ref{7.4}), the Schwinger function $s_n$ (\ref{1279}) can be represented
as
\mar{7.8}\ben
&& s_n(y_1-y_2,\ldots,y_{n-1}-y_n)= \label{7.8}\\
&& \qquad \int\exp[p^j_0(y^0_j-y^0_{j+1})
-i\op\sum^3_{k=1}p^j_k (y^k_j- y^k_{j+1})]w_n^F(p^1,\ldots,p^n)
d_4p^1\cdots d_4p^{n-1}, \nonumber
\een
where $w^F_n\in S'(\ol\Bbb R_{n+})$ is the Fourier transform
of the Wightman function $w_n$, seen as an element of $S'(\Bbb R_{n+})$
of support in the subset $p^i_0\geq 0$.
The formula (\ref{7.8}) enables one to
restore the Wightman functions on the
Minkowski from the Schwinger functions on the Euclidean space
\cite{sim74,zin}.

\end{document}